\documentclass[preprint,12pt, a4paper]{elsarticle}
\usepackage{hyperref}
\UseRawInputEncoding
\usepackage{listings}
\usepackage{amsmath}
\usepackage{amssymb}
\usepackage{comment} 
\usepackage{color}
\usepackage{dcolumn}
\usepackage{bm}
\definecolor{codegreen}{rgb}{0,0.6,0}
\definecolor{codegray}{rgb}{0.5,0.5,0.5}
\definecolor{codepurple}{rgb}{0.58,0,0.82}
\definecolor{backcolour}{rgb}{0.95,0.95,0.92}

\lstdefinestyle{mystyle}{
    backgroundcolor=\color{backcolour},   
    commentstyle=\color{codegreen},
    keywordstyle=\color{magenta},
    numberstyle=\tiny\color{codegray},
    stringstyle=\color{codepurple},
    basicstyle=\ttfamily\footnotesize,
    breakatwhitespace=false,         
    breaklines=true,                 
    captionpos=b,                    
    keepspaces=true,                 
    numbers=left,                    
    numbersep=5pt,                  
    showspaces=false,                
    showstringspaces=false,
    showtabs=false,                  
    tabsize=2
}

\newcommand{\e}{\text{e}} 
\newcommand{\av}[1]{\left\langle #1 \right\rangle} 
\usepackage{lineno}

\usepackage{float}
\restylefloat{table}

\journal{SoftwareX}

\begin{document}

\begin{frontmatter}


\title{
GomalizingFlow.jl: 
A Julia package for Flow-based sampling algorithm
for lattice field theory
}


\author{Akio Tomiya}
\address{{Faculty of Technology and Science}, {International Professional University of Technology}, {{3-3-1}, {Umeda, Kita-ku, Osaka}, {530-0001}, {Osaka}, {Japan}}}
\author{Satoshi Terasaki}%
\address{%
 {AtelierArith}, {980-0004}, {Miyagi}, {Japan}}%

\begin{abstract}
GomalizingFlow.jl: is a package to generate configurations for quantum field theory on the lattice using the flow-based sampling algorithm in Julia programming language. 
This software serves two main purposes: to accelerate research of lattice QCD with machine learning with easy prototyping, and to provide an independent implementation to an existing public Jupyter notebook in Python/PyTorch.
GomalizingFlow.jl implements,
the flow based sampling algorithm, namely, RealNVP and Metropolis-Hastings test for two dimension and three dimensional scalar field, which can be switched by a parameter file. 
HMC for that theory also implemented for comparison.
This package has a Docker image, which reduces the effort for environment construction.
This code works on both the CPU and the NVIDIA GPU. 
\end{abstract}


\end{frontmatter}

\section*{Required Metadata}
\label{}

\section*{Current code version}
\label{}

\begin{table}[H]
\begin{tabular}{|l|p{6.5cm}|p{6.5cm}|}
\hline
\textbf{Nr.} & \textbf{Code metadata description} & \textbf{Please fill in this column} \\
\hline
C1 & Current code version & v1.0.0 
\\
\hline
C2 & Permanent link to code/repository used for this code version & 
\url{
https://github.com/AtelierArith/GomalizingFlow.jl
}
\\
\hline

C3 & Code Ocean compute capsule &
To be uploaded. Docker image available \\
\hline
C4 & Legal Code License   & 
MIT\\
\hline
C5 & Code versioning system used & Git\\
\hline
C6 & Software code languages, tools, and services used & 
Julia\\
\hline
C7 & Compilation requirements, operating environments \& dependencies & Julia 1.8.2, Flux.jl v0.13.5, (Docker image is provided)\\
\hline
C8 & If available Link to developer documentation/manual & 
\url{https://github.com/AtelierArith/GomalizingFlow.jl/blob/main/README.md}
\\ 
\hline
C9 & Support email for questions &
akio@yukawa.kyoto-u.ac.jp \\
\hline
\end{tabular}
\caption{Code metadata (mandatory)}
\label{} 
\end{table}

\clearpage

\section{Motivation and significance}
\label{}
A lattice gauge theory with Markov chain Monte-Carlo calculations provides quantitative information on quantum field theory
and particle physics \cite{Frezzotti:2000nk, Capitani:1998mq, ParticleDataGroup:2000nwm, Aoki:2021kgd}.
Regularized quantum field theory is defined on discretized spacetime (lattice), and the discretization makes degrees of freedom finite.
This enables us to perform numerical calculations with Markov chain Monte-Carlo on supercomputers.
Quantum field theory on a lattice has been applied mainly for QCD (Quantum Chromo-dynamics), which is a fundamental theory for the inside of nuclei \cite{Gattringer:2010zz}. 
Lattice calculations of QCD (Lattice QCD) have succeeded in reproducing the hadron spectrum \cite{Durr:2008zz}, and enable us to access non-perturbative and quantitative information which is difficult to obtain with other methods \cite{HotQCD:2014kol, Borsanyi:2010cj, Aoyama:2020ynm}.

The ultraviolet cutoff is necessary to define a finite theory.
However, the ultraviolet cutoff, namely finite lattice spacing, bring unwanted artifacts into the theory.
Finer lattice spacing gives smaller lattice artifacts, but it causes long autocorrelation among the Monte Carlo samples, which makes Monte Carlo calculations inefficient because the random samples produced are correlated, which is called critical slowing down \cite{Schaefer:2010hu}. 
To make Monte-Calo samples in good quality in a short time, a new algorithm with shorter autocorrelation is demanded because the autocorrelation time depends on an algorithm. 

The trivializing map has been investigated to resolve the critical slowing down in lattice QCD calculations \cite{Luscher:2009eq}. 
Trivializing maps is a change of variables in the (path-)integral between a physical theory to a trivial theory, which is easy to calculate.
A concrete example of the trivializing map called the Nicoli map has been realized in a supersymmetric field theory in the first \cite{Nicolai:1979nr}. 
For non-super symmetric theory, M. Luscher has shown that the gradient flow with Jacobian is an approximate trivializing map, 
but his original idea has not worked well because of the numerical cost.

Machine learning algorithms to generate configurations for lattice field theory are widely investigated \cite{Boyda:2022nmh}.
The first trial has been performed with the Boltzmann machine \cite{Tanaka:2017niz}, 
and others such as GAN \cite{Pawlowski:2018qxs,Zhou:2020yna} have been tried.
The gauge covariant neural network \cite{Tomiya:2021ywc} is applied to perform simulations for four-dimensional QCD.
A neural network parameterized leapfrog integrator enhances topology change in simulations \cite{Foreman:2021rhs}.
The normalizing flow \cite{
Albergo:2019eim,
Kanwar:2020xzo,
Boyda:2020hsi,
Albergo:2021bna,
Abbott:2022zhs,
Abbott:2022hkm,
Albergo:2021vyo,
DelDebbio:2021qwf,
Foreman:2021ljl} has recently been extensively investigated, and we discuss this further later.
Generative models are machine learning architecture for image generation, and which has been investigated  to generate good quality images.
Images are two-dimensional data with certain structures, and it seems that it can be applied to generate field configurations.

The flow based sampling algorithm, based on Real-NVP (Non-volume preserving map), which is a generative model, and is an exact algorithm
of Monte-Carlo configuration generation for quantum field theories associated with the Metropolis--Hastings test.
It has the potential to improve critical slowing \cite{Albergo:2019eim, Kanwar:2020xzo}, which has been extended to system with SU(N) links \cite{Boyda:2020hsi} and with fermions \cite{Albergo:2021bna, Abbott:2022zhs, Abbott:2022hkm}.
This can be regarded as a realization of a trivializing map {\it a la} M. Luscher using a neural network.

One of the purposes of quantum field theory is to calculate the expectation value in the following form,
\begin{align}
\av{O} \equiv \int \mathcal{D}\phi\;
P^\text{(QFT)}_{ \{L,m^2,\lambda\} }[\phi]\;
O[\phi], \label{eq:def_expt}
\end{align}
where 
\begin{align}
\mathcal{D}\phi\; = \prod_{n=n_o}^{L^d} d\phi_n
\end{align}
and $n_o = \underbrace{(1,1,\cdots)}_{d}$, the origin\footnote{
In Julia programming language, index is started from $1$ mostly.
We follow this notation.
}, $d$ is the dimensionality, and $L$ is a size of the system.
And $P^\text{(QFT)}_{ \{L,m^2,\lambda\} }[\phi]$ is a probability density of the system,
\begin{align}
P^\text{(QFT)}_{ \{L,m^2,\lambda\} }[\phi]
&=
\frac{1}{Z}\e^{-S^\text{(QFT)}[\phi]}, \label{eq:def_prob}
\end{align}
and $Z$ normalizes the expectation values as $\av{1} =1$.
The action is,
\begin{align}
S^\text{(QFT)}[\phi] =- \sum_{n=n_o}^{L^d} \phi(n_o) \partial^2 \phi(n_o) + 
\sum_{n=n_o}^{L^d}
V[\phi](n).
\end{align}
where $\partial^2 \phi(n)
=\sum_\mu [\phi(n+\hat\mu)+\phi(n-\hat\mu)-2\phi(n)]
$ is a discretized Laplacian and,
\begin{align}
V[\phi](n) = {{m}^2}\phi^2(n) 
+\lambda \phi^4(n),
\end{align}
is a potential term.

To calculate the integral Eq.  \eqref{eq:def_expt}, we use Markov Chain-Monte Carlo.
We generate a sequence of configurations,
\begin{align}
\phi^{(1)} \to
\phi^{(2)} \to
\phi^{(3)} \to
\phi^{(4)} \to
\cdots \to
\phi^{(N_{\rm conf})},
\end{align}
where $\phi^{(k)}$ is a configuration sampled from Eq. \eqref{eq:def_prob} and ${N_{\rm conf}}$ is the number of samples.
Typically, HMC (Hybrid / Hamiltonian Monte Carlo) has been used \cite{Duane:1987de, Clark:2006fx}.
We can calculate the expectation value through
\begin{align}
\av{O} = \lim_{N_{\rm conf} \to \infty} \frac{1}{N_{\rm conf}}
\sum_{k=1}^{N_{\rm conf}}
O[\phi^{(k)}].
\end{align}
In practice, we cannot take $N_{\rm conf}$ infinity, the expectation value is suffered from the statistical error,
\begin{align}
\av{O} = \frac{1}{N_{\rm conf}}
\sum_{k=1}^{N_{\rm conf}}
O[\phi^{(k)}]
+
\mathcal{O}\left(
\frac{1}{\sqrt{N_{\rm indep}}}
\right)
\end{align}
where $N_{\rm indep} $ is the number of independent samples defined by
\begin{align}
N_{\rm indep} = \frac{N_{\rm conf} }{2 \tau_{ac}}
\end{align}
and $\tau_{ac}$ is the autocorrelation time \cite{Luscher:2004pav}.
In some parameter regime, typically critical regime, $\tau_{ac}$ grows and the number of independent configurations $N_{\rm indep}$ becomes small, which is called the critical slowing down.
The autocorrelation time $\tau_{ac}$ depends on an algorithm, and here we implement the flow based sampling algorithm to
reduce the autocorrelation time.

The flow-based sampling algorithm utilizes three probability distributions.
First is, the target distribution.
We want to draw a sample from the target as
\begin{align}
\phi \sim P^\text{(QFT)}_{ \{L,m^2,\lambda\} }[\phi].
\end{align}
where ``$\sim$'' represents ``sampling'' from a probability distribution.

Second one is, a trivial distribution (a prior distribution in terms of machine learning),
\begin{align}
\varphi \sim P^\text{(pri)}_{\vartheta}[\varphi],
\end{align}
which has a set of parameters $\vartheta$, this is a fixed parameter of the calculations (hyperparameter).
This probability distribution should be point-wise independent distribution and this can be regarded as
a physical distribution at infinite temperature for spin system, or, $\beta = 0$
distribution for lattice gauge systems. 
In terms of quantum field theory, one can regard a probability distribution for a field without a kinetic term.

We apply a parameterized transformation (like a cooling process),
\begin{align}
\phi = \mathcal{F}^{-1}_\theta[\varphi],
\end{align}
where $\theta$ is a set of parameters on the map and
such that,
\begin{align}
P^\text{(QFT)}_{ \{L,m^2,\lambda\} }[\phi]
\approx
P^\text{(ML)}_\Theta[\phi] 
=
P^\text{(pri)}_{\vartheta}[\varphi]
\Bigg| \frac{\partial \mathcal{F}^{-1}_\theta[\varphi]}{\partial \varphi} \Bigg|,
\end{align}
and  $\Theta = \theta \cup \vartheta$ and
\begin{align}
\Bigg| \frac{\partial \mathcal{F}^{-1}_\theta[\varphi]}{\partial \varphi} \Bigg|,
\end{align}
is the Jacobian for the transformation.
The quality of this map is measured by (shifted and reversed) Kullback–Leibler divergence,
\begin{align}
\int \mathcal{D}\phi \;
P^\text{(ML)}_\Theta[\phi] \ln P^\text{(ML)}_\Theta[\phi] / P^\text{(QFT)}_{ \{L,m^2,\lambda\} }[\phi]- \ln Z,
\end{align} 
and this works as a variational energy of approximation.
By tuning parameters $\theta$ to mimize Kullback–Leibler divergence, 
we can realize approximated (un) trivializing map.

\begin{figure}[t] 
\centering 
\includegraphics[width=1.0\textwidth]{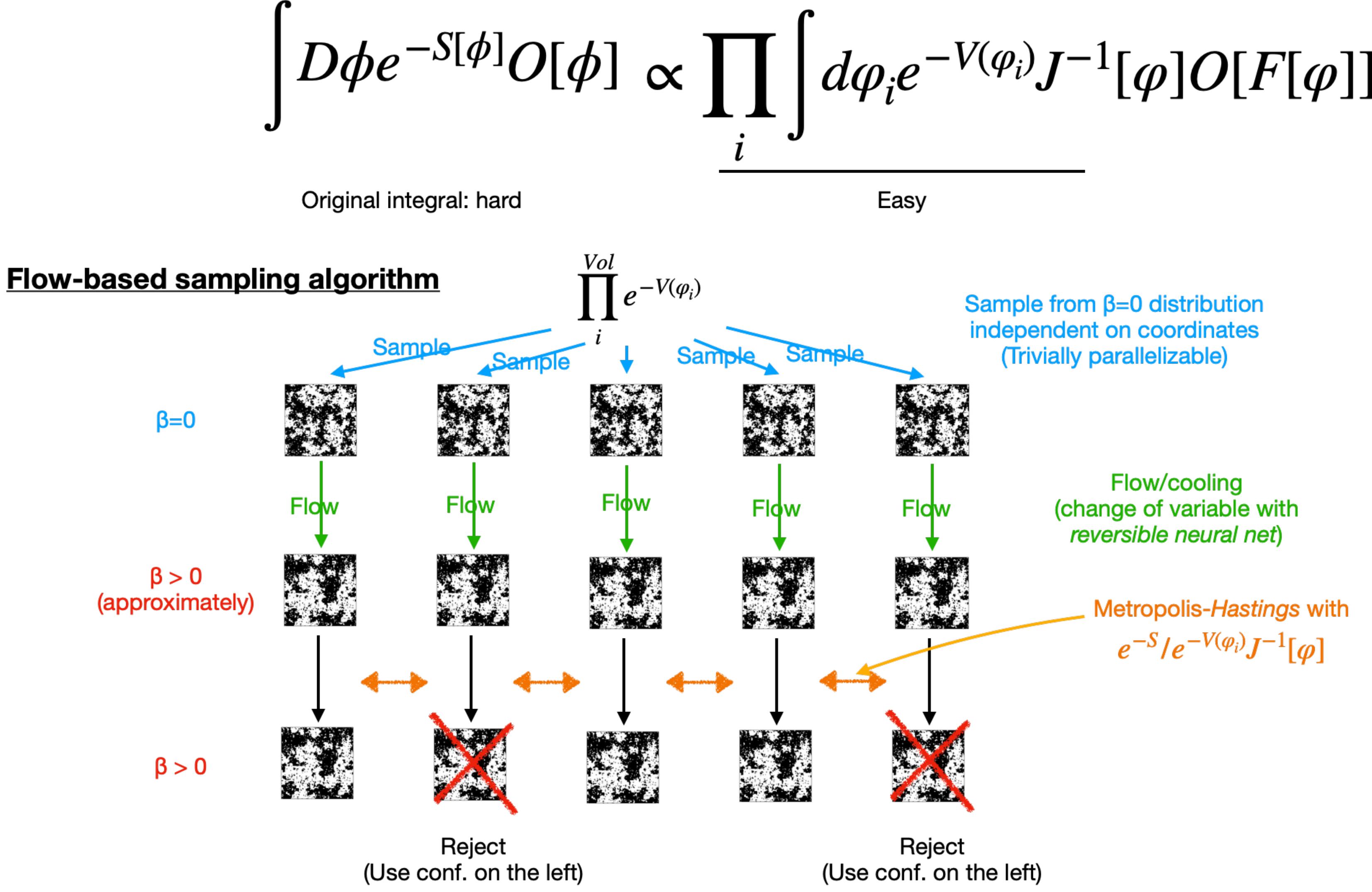}
\caption{Schematic picture for the flow based sampling algorithm.
Here we use $\beta$ to indicate triviality of the system, which is a coefficient of the kinetic term in our current application (please see the main text in details).
\label{schematic}
} 
\end{figure}

Fig. \ref{schematic} is a schematic overview for the flow-based sampling algorithm.
First, it samples from a configuration with point-wise independent distribution (in terms of 
gauge theory, it is sampling from $\beta=0$).
Next, we apply the flow transformation $\mathcal{F}^{-1}_\theta$, un-trivializing map, written by a neural network.
This makes correlations between sites.
Output from the network can be regarded as a sample from $\beta >0$ approximately.
Finally, with the Metropolis-Hastings test with the Jacobian, we obtain an ensemble which obeys the 
correct distribution ${\e}^{-S}/Z$.
Training also follows the process described above.

Here, we describe how we can realize the map $\mathcal{F}_\theta$ (trivializing map) and 
$\mathcal{F}^{-1}_\theta$ (untrivializing map) in terms of neural networks.
The structure of the affine coupling layer is essential for a tractable Jacobian.
We employ checker-board decomposition to get tractable Jacobian as in the original work \cite{Albergo:2019eim}. 
We alternately apply small and reversible trivializing maps, which are realized using a neural network, on even and odd sites. 
By composing these maps, we approximately realize a map between the target distribution to the trivial distribution and one for the opposite direction.
The affine coupling layer is realized as follows.

We consider the affine coupling layer on odd sites first. 
We denote $\phi$ and $\varphi$ as nontrivial and trivial side variables, respectively. 
The affine coupling layer (scaling and shift transformation for field variable) is defined as,
\begin{align}
\begin{cases}
\varphi_e &= \e^{s[\phi_o]} \phi_e + t[\phi_o],\\
\varphi_o &=  \phi_o 
\end{cases}
\end{align}
where $s$ and $t$ are output of a neural network,
\begin{align}
T^{o}_{\theta}:
\phi_o
\mapsto
\begin{pmatrix}
s\\t
\end{pmatrix}
=
\begin{pmatrix}
s\\t
\end{pmatrix} [\phi_o] = 
\begin{pmatrix}
s\\t
\end{pmatrix} (\phi_1,\phi_3,\cdots)
\end{align}
and the arguments are field values on odd sites for a configuration.
We remark that this transformation is bijective.
Transformations for even sites $T^{e}_{\theta}$ are defined in a similar manner.
In this work, this map is realized using a two- or three-dimensional convolutional neural network.
We can realize an approximate trivializing map by applying this map with different weights repeatedly,
\begin{align}
\mathcal{F}_{\theta}[\phi ] = 
T^{o}_{\theta_{N_{\rm tri}}}
\circ
T^{e}_{\theta_{N_{\rm tri}-1}}
\circ
\cdots
\circ
T^{e}_{\theta_{2}}
\circ
T^{o}_{\theta_{1}}
[\phi]. 
\end{align}
where $\theta = \theta_1 \cup \theta_2 \cup
\cdots \cup {\theta_{N_{\rm tri}}} $ and
$\theta_i$ is a set of weights in each neural network, and
 ${N_{\rm tri}} $ is the number of affine coupling layers.
This map $\mathcal{F}_{\theta}[\varphi ] $ is also bijective, since $T$ is bijective.
Thus, we can obtain an untrivializing map $\mathcal{F}^{-1}_{\theta}$.
Due to the even odd structure of the transformation, the Jacobian calculation is $O(N)$ \cite{Albergo:2019eim}.
We remark that, (un-)tirivializing procedure is analogous to integration steps in the gradient flow \cite{Tomiya:2021ywc}.

This package is implemented by Julia, which is a new scientific programming language \cite{https://doi.org/10.48550/arxiv.1209.5145}.
Typically Julia is as fast as C and Fortran \cite{juliabenchmark}, and is fit to calculations in the lattice QCD community.
The flow based sampling algorithm has been implemented in Python/PyTorch \cite{Albergo:2021vyo},
and this package provides another implementation.

\section{Software description}
\label{}

\subsection{Software Architecture}
\label{}

\begin{figure}[t] 
\centering 
\includegraphics[width=1.0\textwidth]{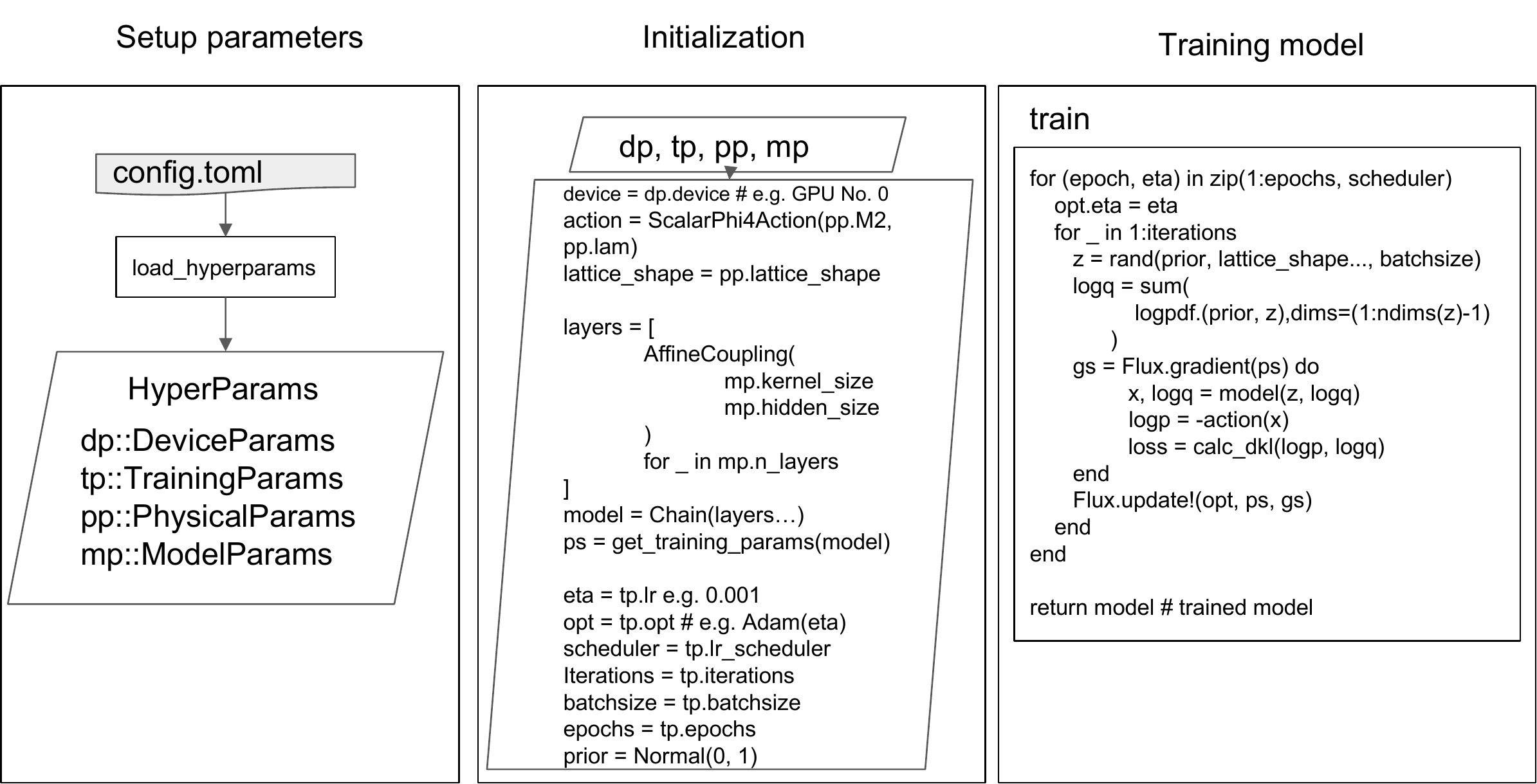}
\caption{An illustration of code architecture
} 
\label{arc}
\end{figure}

This package works as follows (Fig. \ref{arc}).
First, a parameter configuration file ({\it e.g.} ``cfgs/example2d.toml'') is loaded.
The parameters in the file are classified into four groups.
$DeviceParams$ ($dp$) specifies what kind of acceleration device is used (-1 for no accelerator).
$TrainingParams$ ($tp$)
controls the training procedure ({\it, e.g.} epoch, batch size, optimizer). 
$PhysicalParams$ ($pp$) should contain parameters of the target system (mass, coupling, and system size).
$ModelParams$ ($mp$) is a set of parameters for a neural network in the flow model.
These parameters are loaded through $load\_hyperparams$ function.

The code constructs objects according to the loaded hyperparameters and physical parameters in the memory.
The objects is for example,
action of the target theory, neural network model, an optimizer, and setting of training like,
the number of epochs, batch size, and scheduling of learning rate ($lr$).

The simulation is performed with parameters with the action defined in $src/actions.jl$.
Jacobian can be calculated from the output of the neural network.

\subsection{Software Functionalities}
\label{}
In the current version, the flow based sampling algorithm for 2 and 3 dimensional scalar field theory can be used in a Docker environment, which is explained in the following section.
HMC for these theories are implemented to compare results. HMC uses the same struct for parameters.

Jupyter lab environment for analysis, {\it e.g.} the condensate, zero-momentum two point functions, calculation of the autocorrelation time, are available in $playground/notebook/julia/analysis\_tool.md$. 
This file is a markdown file but will be automatically converted to an ipynb file via jupytext.

\label{}

\section{Illustrative Examples}
\label{}

The use of Docker images is the easiest way to use this package\footnote{
To use Docker with NVIDIA GPU, we recommended to follow NVIDIA provided procedure in \url{https://docs.nvidia.com/datacenter/cloud-native/container-toolkit/install-guide.html\#docker}.
}.
In the repository, one can execute the training, with commands in Listing \ref{com:execute}.
A file $cfgs/example2d.toml$ is a parameter file, which is explained below.
\begin{lstlisting}[caption={Command to start training},label=com:execute]
$ make
$ docker-compose run --rm julia julia begin_training.jl cfgs/example2d.toml
\end{lstlisting}
Then the training for the two dimensional scalar field will be started.
The results are saved in the $results$ directory.

We provide example files for two-dimensional scalar field theory and three-dimensional scalar field theory for broken and symmetric phases.
Hyper-parameters for training and parameters for the physical system can be written as 
Listing \ref{com:setting_file}.
This performs training with 200 epochs and a prior distribution a Gaussian $N(0,1)$.
A neural network has 16 layers with hidden size, $8 \times 8$.
The training parameter can be understood by comments that started on line 14 of the list.

\begin{lstlisting}[caption={example2d.toml},label=com:setting_file]
[config]
version = "0.1.0"

[device]
# setup DeviceParams for example
# device_id = -1 # <--- train with CPU
# device_id = 0 # <--- train with GPU its Device ID is 0
# device_id = 1 # <--- train with GPU its Device ID is 1
device_id = -1


# setup TrainingParams
#
# for _ in 1:epochs
#    for _ in 1:iterations
#       # extract batchsize data
#       z = rand(prior, lattice_shape..., batchsize)
#       gs = Flux.gradient(ps) do
#                do something
#       end
#       Flux.Optimise.update!(opt(base_lr), ps, gs)
#    end
# end
[training]
seed = 12345
batchsize = 64
epochs = 200
iterations = 100
base_lr = 0.001
opt = "Adam"
prior = "Normal{Float32}(0.f0, 1.f0)"
lr_scheduler = ""
pretrained = ""

# setup PhysicalParams
#
# lattice_shape = (L, L) if Nd = 2
# lattice_shape = (L, L, L) if Nd = 3
# action = ScalarPhi4Action(M2, lam)
[physical]
L = 8
Nd = 2
M2 = -4.0 # m^2
lam = 8.0 # λ


# setup ModelParams
#
[model]
seed = 2021
n_layers = 16
hidden_sizes = [8, 8]
kernel_size = 3
inC = 1
outC = 2
use_final_tanh = true
use_bn = false
\end{lstlisting}

\section{Impact}
\label{}

This software enables us to perform extensive research with the flow based sampling algorithm in two dimension and three dimension on CPU and GPU, which is preferred since modern computer clusters and supercomputers have GPU accelerator.
More and more work with the flow-based sampling algorithm is being published, and most of them are based on a Python code \cite{Albergo:2021vyo}.
We provide a new code in the Julia language with Docker environment. 
Thanks to the Julia environment, this package can be run on a large-scale supercomputer or a cluster with GPU if the machine supports Julia
language.

This package also has the advantage of being used for prototyping.
If one wants to change the theory, one can achieve it by changing $src/actions.jl$.
Differentiation can be done by Flux.jl/Zygote.jl automatically as is explained above.

\section{Conclusions}
\label{}

We implement a flow-based sampling algorithm for two or three-dimensional lattice scalar field theory in Julia programming language.
This package extensively uses automatic differentiation in Flux.jl/Zygote.jl to calculate Jacobian in the flow based sampling algorithm.
In addition, this package provides, HMC, analysis tools for autocorrelation and ESS (effective sample size) calculation with Jupyter notebook.
A Docker image is provided, which enables us to perform simulations without effort on environment construction
and this package works not only on the CPU but also on the NVIDIA GPU.
Hyperparameters and parameters including dimensionality can be specified in the a parameter file.

This package pave a way of research directions not only the reduction of autocorrelation time \cite{Albergo:2019eim,
Kanwar:2020xzo,Albergo:2021bna},
scaling study \cite{DelDebbio:2021qwf}, and investigation with sign problem \cite{Rodekamp:2022xpf}.

\section{Conflict of Interest}
There are no known conflicts of interest associated with this publication and there has not been significant financial support for this work that could have influenced its outcome.

\section*{Acknowledgements}
\label{}
Authors thank to a public notebook \cite{Albergo:2021vyo}, and we refer the code for the implementation.
This work of AT was  supported by JSPS KAKENHI Grant Number 
JP20K14479, 
JP22H05112, 
JP22H05111, and 
JP22K03539. 
This work was supported by MEXT as ``Program for Promoting Researches on the Supercomputer Fugaku'' (Simulation for basic science: approaching the new quantum era; Grant Number JPMXP1020230411).
This work was supported by MEXT as ``Program for Promoting Researches on the Supercomputer Fugaku'' (Search for physics beyond the standard model using large-scale lattice QCD simulation and development of AI technology toward next-generation lattice QCD; Grant Number JPMXP1020230409).

\bibliographystyle{elsarticle-num} 



\bibliography{ref}
\end{document}